\documentstyle[12pt]{article}
\input epsf
\newcommand{\nc}{\newcommand}
\nc{\postscript}[2] 
{\setlength{\epsfxsize}{#2\hsize}\centerline{\epsfbox{#1}}}
\nc{\bg}{B. Grz\c{a}dkowski}
\nc{\non}{\nonumber}
\nc{\barx}{\bar{x}}\nc{\pbarn}{\;\hbox {pb}}\nc{\fbarn}{\;\hbox {fb}}
\nc{\hc}{\hbox {h.c.}} \nc{\re}{\hbox {Re}} \def\im{{\rm Im}}
\nc{\mev}{\hbox {MeV}} \nc{\gev}{\;\hbox {GeV}}
\def\gesim{\lower0.5ex\hbox{$\:\buildrel >\over\sim\:$}} 
\def\lesim{\lower0.5ex\hbox{$\:\buildrel <\over\sim\:$}} 
\nc{\prd}[3]{{\it Phys.\ Rev.}\ {{\bf D{#1}} (#2), #3}}
\nc{\prl}[3]{{\it Phys.\ Rev.\ Lett.}\ {{\bf {#1}} (#2), #3}}
\nc{\plb}[3]{{\it Phys.\ Lett.}\ {{\bf B{#1}} (#2), #3}}
\nc{\npb}[3]{{\it Nucl.\ Phys.}\ {{\bf B{#1}} (#2), #3}}
\nc{\ptp}[3]{{\it Prog.\ Theor.\ Phys.}\ {{\bf {#1}} (#2), #3}}
\nc{\zfp}[3]{{\it Z.\ Phys.}\ {{\bf C{#1}} (#2), #3}}
\nc{\mpla}[3]{{\it Mod.\ Phys.\ Lett.}\ {{\bf A{#1}} (#2), #3}}
\nc{\rmp}[3]{{\it Rev.\ Mod.\ Phys.}\ {{\bf {#1}} (#2), #3}}
\nc{\ijmpa}[3]{{\it Int.\ J.\ of\ Mod.\ Phys.}\
               {{\bf A{#1}} (#2), #3}}
\nc{\app}[3]{{\it Acta\ Phys.\ Polon.}\ {{\bf B{#1}} (#2), #3}}

\nc{\ttbar}{t\bar{t}}         \nc{\bbbar}{b\bar{b}}
\nc{\tanb}{\tan \beta}        \nc{\twbdec}{t\to W^+ b}
\nc{\tbwbdec}{\bar{t}\to W^- \bar{b}}
\nc{\epem}{e^+e^-}            \nc{\eett}{\epem \to \ttbar}
\nc{\sigeett}{\sigma_{e\bar{e}\to\ttbar}}
\nc{\wpwm}{W^+W^-}            \nc{\tbar}{\bar{t}}
\nc{\bbar}{\bar{b}}           \nc{\wpp}{W^+}
\nc{\mt}{m_t}    \nc{\mts}{m_t^2}   \nc{\mw}{M_W}    \nc{\mws}{M_W^2}
\nc{\mz}{M_Z}    \nc{\mzs}{M_Z^2}
\nc{\ttbardec}{\ttbar \to W^+W^-\bbbar}
\nc{\wwbb}{W^+W^-\bbbar}      \nc{\sm}{SM}
\nc{\cw}{\cos\theta_W}        \nc{\sw}{\sin\theta_W}
\nc{\sws}{\sin^2\theta_W}     \nc{\sig}{\sigma_{tot}}
\nc{\lp}{\ell^+}              \nc{\lm}{\ell^-}
\nc{\epsl}{\epsilon_L}        \nc{\cp}{C\!P}

\nc{\splus}{s_+}       \nc{\smin}{s_-}        \nc{\eps}{\epsilon}
\nc{\psp}{Ps_+}        \nc{\psm}{Ps_-}        \nc{\lsp}{ls_+}
\nc{\lsm}{ls_-}        \nc{\sss}{s_+s_-}      \nc{\m}{m_t}
\nc{\mq}{m_t^2}        \nc{\mr}{\frac{1}{\m}} \nc{\av}{A_{\gamma}}
\nc{\bv}{B_{\gamma}}   \nc{\az}{A_Z}          \nc{\bz}{B_Z}
\nc{\avs}{A_{\gamma}^2}\nc{\azs}{A_Z^2}       \nc{\bzs}{B_Z^2}
\nc{\dav}{\delta \! A_{\gamma}}   \nc{\dbv}{\delta \! B_{\gamma}}
\nc{\dcv}{\delta C_{\gamma}}      \nc{\ddv}{\delta \! D_{\gamma}}
\nc{\daz}{\delta \! A_Z}          \nc{\dbz}{\delta \! B_Z}
\nc{\dcz}{\delta C_Z}             \nc{\ddz}{\delta \! D_Z}
\nc{\dev}{\delta \! E_{\gamma}}   \nc{\dez}{\delta \! E_Z}
\nc{\dfv}{\delta \! F_{\gamma}}   \nc{\dfz}{\delta \! F_Z}
\nc{\rdav}{{\rm Re}(\delta \! A_{\gamma}) \:}
\nc{\rdbv}{{\rm Re}(\delta \! B_{\gamma}) \:}
\nc{\rdcv}{{\rm Re}(\delta C_{\gamma}) \:}
\nc{\rddv}{{\rm Re}(\delta \! D_{\gamma}) \:}
\nc{\rdaz}{{\rm Re}(\delta \! A_Z) \:}
\nc{\rdbz}{{\rm Re}(\delta \! B_Z) \:}
\nc{\rdcz}{{\rm Re}(\delta C_Z) \:}
\nc{\rddz}{{\rm Re}(\delta \! D_Z) \:}
\nc{\idav}{{\rm Im}(\delta \! A_{\gamma}) \:}
\nc{\idbv}{{\rm Im}(\delta \! B_{\gamma}) \:}
\nc{\idcv}{{\rm Im}(\delta C_{\gamma}) \:}
\nc{\iddv}{{\rm Im}(\delta \! D_{\gamma}) \:}
\nc{\idaz}{{\rm Im}(\delta \! A_Z) \:}
\nc{\idbz}{{\rm Im}(\delta \! B_Z) \:}
\nc{\idcz}{{\rm Im}(\delta C_Z) \:}
\nc{\iddz}{{\rm Im}(\delta \! D_Z) \:}
\nc{\cz}{(1+v_e^2)d\:\!'^2}         \nc{\ci}{v_ed\:\!'}
\nc{\ccz}{v_ed\:\!'^2}              \nc{\cci}{d\:\!'}
\nc{\lspace}{\;\;\;\;\;\;\;\;\;\;}  \nc{\llspace}{\lspace \lspace}
\nc{\ra}{\rightarrow} 

\nc{\beq}{\begin{equation}}   \nc{\eeq}{\end{equation}}
\nc{\bea}{\begin{eqnarray}}   \nc{\eea}{\end{eqnarray}}
\nc{\baa}{\begin{array}}      \nc{\eaa}{\end{array}}
\nc{\bit}{\begin{itemize}}    \nc{\eit}{\end{itemize}}
\nc{\ben}{\begin{enumerate}}  \nc{\een}{\end{enumerate}}
\nc{\bce}{\begin{center}}     \nc{\ece}{\end{center}}
\begin{document}
\pagestyle{empty} \setlength{\footskip}{2.0cm}
\setlength{\oddsidemargin}{0.5cm} \setlength{\evensidemargin}{0.5cm}
\renewcommand{\thepage}{-- \arabic{page} --}
\def\mib#1{\mbox{\boldmath $#1$}}
\def\bra#1{\langle #1 |}      \def\ket#1{|#1\rangle}
\def\vev#1{\langle #1\rangle} \def\dps{\displaystyle}
   \def\thebibliography#1{\centerline{REFERENCES}
     \list{[\arabic{enumi}]}{\settowidth\labelwidth{[#1]}\leftmargin
     \labelwidth\advance\leftmargin\labelsep\usecounter{enumi}}
     \def\newblock{\hskip .11em plus .33em minus -.07em}\sloppy
     \clubpenalty4000\widowpenalty4000\sfcode`\.=1000\relax}\let
     \endthebibliography=\endlist
   \def\sec#1{\addtocounter{section}{1}\section*{\hspace*{-0.72cm}
     \normalsize\bf\arabic{section}.$\;$#1}\vspace*{-0.3cm}}
\vspace*{-1.6cm}\noindent
\hspace*{11.cm}IFT-10-97\\
\hspace*{11.cm}TOKUSHIMA 97-01\\
\hspace*{11.cm}(hep-ph/9710358)\\

\vspace*{.5cm}

\begin{center}
{\large\bf Effects of Non-Standard Interactions for the Energy}

\vskip 0.1cm
{\large\bf Spectrum of Secondary Leptons in $\!\!\mib{e}^+\mib{e}^-
\!\to\mib{t}\bar{\mib{t}}$} 
\end{center}

\vspace*{1cm}
\begin{center}
\renewcommand{\thefootnote}{\alph{footnote})}
{\sc Longin BRZEZI\'NSKI}$\,$\footnote{E-mail address:
\tt longin.brzezinski@fuw.edu.pl},
{\sc Bohdan GRZ\c{A}DKOWSKI}$\,$\footnote{E-mail address:
\tt bohdan.grzadkowski@fuw.edu.pl}

\vspace*{.5cm}
{\sl Institute of Theoretical Physics,\ Warsaw 
University}\\
{\sl Ho\.za 69, PL-00-681 Warsaw, POLAND} 

\vspace*{1.cm}
{\sc Zenr\=o HIOKI}$\,$\footnote{E-mail address:
\tt hioki@ias.tokushima-u.ac.jp}

\vspace*{.5cm}
{\sl Institute of Theoretical Physics,\ University of Tokushima}\\
{\sl Tokushima 770-8502, JAPAN}
\end{center}

\vspace*{1.5cm}
\centerline{ABSTRACT}

\vspace*{0.4cm}
\baselineskip=20pt plus 0.1pt minus 0.1pt
The process of top-quark pair production followed by semileptonic
decays at future high-energy $e^+ e^-$ linear colliders is
investigated as a possible test of physics beyond the Standard Model.
Assuming the most general non-standard forms for $\gamma \ttbar$, $Z
\ttbar$ and $Wtb$ couplings, the energy spectrum of the single lepton
$\ell^\pm$ and the energy correlation of $\lp$ and $\lm$ emerging
from the process $e^+e^-\!\to t\bar{t}\to \ell^\pm \cdots /\ell^+
\ell^- \cdots $ are calculated. Expected precision of the
non-standard-parameter determination is estimated adopting the
recently-proposed optimal method.

\vfill
\newpage
\renewcommand{\thefootnote}{\sharp\arabic{footnote}}
\pagestyle{plain} \setcounter{footnote}{0}
\baselineskip=21.0pt plus 0.2pt minus 0.1pt

\sec{Introduction}

High-energy $\epem$ linear collider (NLC) can provide a very useful
laboratory to study physics of the top quark. In spite of spectacular
successes of experimental high-energy physics ({\it e.g.} precision
tests of the Standard Model (SM) of electroweak interactions), the
top-quark couplings have not been tested yet. We should not take it
for granted from the beginning that the top-quark interactions obey
the scheme provided by the SM. The aim of this paper is to consider
possible non-standard effects in the energy spectrum of secondary
lepton(s) emerging in the process $\epem \to \ttbar \to \ell^\pm
\cdots/\ell^+\ell^- \cdots$ at NLC. Since the top quark is heavy,
$m_t^{exp}=175.6 \pm 5.5$~GeV \cite{data97}, it decays as a single
quark before forming bound states. Thanks to this property it is
possible to avoid complicated non-perturbative effects brought
through fragmentation processes in contrast to a case of lighter
quarks.

The leptonic energy spectrum has been studied in the existing
literature \cite{CKP}--\cite{GH_plb}. However, non of those articles
assumed the most general form for the interactions of $\gamma\ttbar$,
$Z \ttbar$ and $W tb$. Although our recent papers
\cite{GH_npb,GH_plb} treated consistently non-standard effects in the
production and decay of top quarks at NLC, we focused our discussion
on $\cp$-violating couplings only.\footnote{$\cp$ violation in
    top-quark production at NLC and in its decay has been discussed
    by very many authors, an incomplete list of references which are
    not cited in the text could be found in ref.\cite{cp}.}\  
In this paper we will present a comprehensive analysis taking into
account the most general non-standard couplings with both
$\cp$-violating and $\cp$-conserving terms.\footnote{We treat all the
    other couplings as in the SM since it is well known that they are
    successfully described within the SM.}

The paper is organized as follows. In sec.\ 2 we will briefly
describe a formalism for the energy-spectrum calculation. Section 3
will provide the differential cross section for polarized $\ttbar$
production assuming the most general $\gamma \ttbar$ and $Z\ttbar$
interactions. In sec.\ 4 we will consider the top-quark decay, where
again we use the most general form for non-standard $Wtb$
interactions. Section 5 will contain the derivation of the single and
the double lepton-energy spectrum. Then, in sec. 6, we will discuss
how to measure all the non-standard couplings using the method of
optimal observables \cite{optimalization}. We summarize our results
in sec. 7. There we also compare our results to those in other
numerical analysis \cite{frey}--\cite{schmitt}. In the appendix, in
order to provide readers some more concrete image, we show
contributions of the dimension 6 operators to the form factors in the
$\gamma t\bar{t}/Zt\bar{t}$ and $Wtb$ couplings in the framework of
effective lagrangian approach \cite{bw}.
  
\sec{The lepton-energy spectrum: standard-model results.}
In this section we briefly present the formalism which will be used
in this paper. For completeness we also show some standard-model
results.

We will treat all the fermions except the top quark as massless and
adopt the technique developed by Kawasaki, Shirafuji and Tsai
\cite{technique}. This is a useful method to calculate distributions
of final particles appearing in a  production process of on-shell
particles and their subsequent decays. This technique is applicable
when the narrow-width approximation
$$
\left|\,{1\over{p^2-m^2+im{\mit\Gamma}}}\,\right|^2
\simeq{\pi\over{m{\mit\Gamma}}}\delta(p^2 -m^2)
$$
can be adopted for the decaying intermediate particles. In fact, this
is very well satisfied for both $t$ and $W$ since ${\mit\Gamma}_t
\simeq$ 175$(\mt/\mw)^3\;\mev \ll\mt$ and ${\mit\Gamma}_W=2.07\pm
0.06$ GeV \cite{PDG} $\ll M_W$.

Adopting this method, one can derive the following formula for the
inclusive distribution of the single-lepton $\ell^+$ in the reaction
$\eett$\ \cite{AS}:
\begin{eqnarray}
&&\frac{d^3\sigma}{d^3 p_\ell/(2p_\ell^0)}(\epem \to \ell^+ + \cdots)
\non \\
&&\ \ \ \ \ \ \ \ \ \ \ \ \ \ \ \ \ 
=4\int d{\mit\Omega}_t
\frac{d\sigma}{d{\mit\Omega}_t}(n,0)\frac{1}{{\mit\Gamma}_t}
\frac{d^3{\mit\Gamma}_\ell}{d^3 p_\ell/(2p_\ell^0)}(t\to b\ell^+\nu),
\label{master}
\end{eqnarray}
where ${\mit\Gamma}_\ell$ and ${\mit\Gamma}_t$ are the leptonic and
total widths of {\it unpolarized} top respectively, and
$d\sigma(n,0)/d{\mit\Omega}_t$ is obtained from the angular
distribution of $\ttbar$ with spins $s_+$ and $s_-$ in $\eett$,
$d\sigma(s_+,s_-)/d{\mit\Omega}_t$, by the following replacement:
\begin{equation}
s^\mu_+ \to n^\mu=\left(g^{\mu \nu}-\frac{p_t^\mu p_t^\nu}{\mts}
\right)\frac{\mt}{p_t p_\ell}p_{\ell\,\nu}
\lspace{\rm and}\lspace s_- \to 0.
\label{replacement}
\end{equation}
(Exchanging the roles of $s_+$ and $s_-$ and reversing the sign of
$n^\mu$, we get the distribution of $\ell^-$.)

Following ref.\cite{AS}, let us introduce the rescaled
lepton-energy, $x$, by
$$
x\equiv
\frac{2 E_\ell}{\mt}\left(\frac{1-\beta}{1+\beta}\right)^{1/2},
$$
where $E_\ell$ is the energy of $\ell$ in $\epem$ c.m. frame and
$\beta=\sqrt{1-4\mts/s}$ ($s\equiv(p_{e^+}+p_{e^-})^2)$. We also
define parameters $D^{(0)}_V$, $D^{(0)}_A$ and $D^{(0)}_{V\!\!A}$ as
\begin{eqnarray}
&&D^{(0)}_V=(v_e v_t d-\frac23)^2 +(v_t d)^2, \non\\
&&D^{(0)}_A=(v_e d)^2 +d^2,                   \non\\
&&D^{(0)}_{V\!\!A}=v_e d(v_e v_t d-\frac23) +v_t d^2,
\label{DVDA}
\end{eqnarray}
by using the standard-model neutral-current couplings for $e$ and
$t$: $v_e=-1+4\sin^2\theta_W$ and $v_t=1-(8/3)\sin^2\theta_W$, and a
$Z$-propagator factor
$$
d\equiv\frac{s}{s-M_Z^2}
\frac{1}{16\sin^2\theta_W\cos^2\theta_W}.
$$

Then, the $x$ spectrum is given in terms of these quantities by
\begin{eqnarray}
\frac{1}{B_\ell\sigeett}{\frac{d\sigma}{dx}}^{\!\pm}\equiv
\frac{1}{B_\ell\sigeett}
\frac{d\sigma}{dx}(\epem \to \ell^\pm+\cdots)=f(x)+\eta\: g(x).
\end{eqnarray}
Here $\sigeett\equiv\sigma_{tot}(e^+e^-\!\to\ttbar)$, $B_\ell$ is the
branching ratio for $t\to \ell +\cdots$ ($\simeq 2/9$ for $\ell=e,
\mu$). $f(x)$ and $g(x)$ are functions introduced in ref.\cite{AS}:
\beq
f(x)=\frac{3}{W} \frac{1+\beta}{\beta} \int d \omega \; \omega\:,
\ \ \ 
g(x)=\frac{3}{W} \frac{1+\beta}{\beta} \int d \omega \; \omega
\left[1-\frac{x(1+\beta)}{1-\omega}\right],
\eeq
where 
$$
W\equiv(1-r)^2(1+2r),\ \ r\equiv(M_W/m_t)^2,\ \ 
\omega\equiv(p_t -p_\ell)^2/m_t^2.
$$
$f(x)$ and $g(x)$ satisfy the following normalization conditions:
$$
\int f(x)dx=1 \lspace {\rm and} \lspace \int g(x)dx=0.
$$
The explicit form of $f(x)$ and $g(x)$ could be found in refs.
\cite{AS} and \cite{GH_npb}.
$\eta$ is defined as
$$
\eta\equiv 4\:a_{V\!\!A}D^{(0)}_{V\!\!A},
$$
where
$a_{V\!\!A}\equiv 1/[\,(3-\beta^2)D^{(0)}_V +2\beta^2 D^{(0)}_A\,]$.

Applying the same technique, we get the following energy correlation
of $\lp$ and $\lm$ :
\begin{equation}
\frac{1}{B_\ell^2 \sigeett} \frac{d^2\sigma}{dx\;d\barx}=
S_0 (x,\barx), \label{S0}
\end{equation}
where $x$ and $\barx$ are the rescaled energies of $\ell^+$ and
$\ell^-$ respectively, and
$$
S_0 (x,\barx)=f(x)f(\barx)+\eta\,[\,f(x)g(\barx)+g(x)f(\barx)\,]
+\eta'g(x)g(\barx)\,
$$
with $\eta'$ being defined as
$$
\eta'\equiv
\beta^{-2}a_{V\!\!A}[\,(1+\beta^2)D^{(0)}_V+2\beta^2 D^{(0)}_A\,].
$$
Clearly, the  $(x,\bar{x})$ distribution is symmetric in $x$ and
$\barx$, which is a sign of the standard-model $C\!P$
symmetry.\footnote{The SM requires at least two-loops to generate
    $\cp$-violating energy distributions.}

In the following, we use $M_W=80.43$ GeV, $M_Z=91.1863$ GeV, $m_t=
175.6$ GeV, $\sin^2\theta_W=0.2315\:$\cite{data97}\ and $\sqrt{s}
=$500 GeV. For these inputs, we have
$$
\eta=0.2074,\ \ \ \eta'=1.2720\ \ \ {\rm and}
\ \ \ a_{V\!\!A}=0.7545.
$$
\setlength{\footskip}{2cm}

\sec{Non-standard effects in the production of polarized $\mib{t}
\bar{\mib{t}}$ }

We will assume that all non-standard effects in the production of
$\ttbar$ can be represented by the following corrections to the
photon and $Z$-boson vertices contributing to the $s$-channel
diagrams:
\begin{equation}
{\mit\Gamma}_{vt\bar{t}}^{\mu}=
\frac{g}{2}\,\bar{u}(p_t)\,\Bigl[\,\gamma^\mu \{A_v+\delta\!A_v
-(B_v+\delta\!B_v) \gamma_5 \}
+\frac{(p_t-p_{\bar{t}})^\mu}{2m_t}(\delta C_v-\delta\!D_v\gamma_5)
\,\Bigr]\,v(p_{\bar{t}}),\ \label{ff}
\end{equation}
where $g$ denotes the $SU(2)$ gauge coupling constant, $v=\gamma,Z$,
and
\[
\av=\frac{4}{3}\sw,\ \ \bv=0,\ \ \az=\frac{v_t}{2\cw},\ \ \bz
=\frac{1}{2\cw}.
\]
In addition, contributions to the vertex proportional to $(p_t +
p_{\bar{t}})^\mu$ are also allowed, but their effects vanish in the
limit of zero electron mass.\footnote{These contributions are
    essential for the $U(1)_{\rm EM}$ gauge invariance. We discuss
    this point briefly in the appendix.}\ 
Among the above new form factors, $\delta\!A_{\gamma,Z},
\delta\!B_{\gamma,Z},\delta C_{\gamma,Z}$ and $\delta\!D_{\gamma,Z}$
are parameterizing $C\!P$-conserving and $C\!P$-violating
non-standard interactions, respectively. In the appendix, we show how
they receive contributions from effective operators of dimension 6.

On the other hand, interactions of initial $\epem$ have been assumed
untouched by non-standard interactions:
\begin{itemize}
\item $\gamma e^+ e^-$ vertex
\beq
{\mit\Gamma}^{\mu}_{\gamma e^+e^-} =
-e\:\bar{v}(p_{e^+})\,\gamma^{\mu}\,u(p_{e^-})\,,
\eeq
\item $Z e^+e^-$ vertex
\beq
{\mit\Gamma}^{\mu}_{Z e^+e^-} = \frac{g}{4 \cw}\,
\bar{v}(p_{e^+})\,\gamma^{\mu}(v_e+\gamma_5)\,u(p_{e^-})\,.
\eeq
\end{itemize}

A tedious but straightforward calculation leads to the following
formula for the angular distribution of polarized top-quark pair in
presence of the above non-standard interactions:
\begin{eqnarray}
&&\frac{d\sigma}{d{\mit\Omega}}(e^+e^-\to t(s_+)\bar{t}(s_-))
\non\\
&&=\frac{3\beta\alpha^2}{16 s^3}\:\Bigl[
\:\:D_V\:[\:\{ 4m_t^2s+(lq)^2 \}(1-s_{+}s_{-})+s^2(1+s_{+}s_{-})
\non\\
&&\ \ \ \lspace +2s(ls_{+}\;ls_{-}-Ps_{+}\;Ps_{-})
           +\:2\,lq(ls_{+}\;Ps_{-}-ls_{-}\;Ps_{+})\:]
\non\\
&&\ \ \ +\:D_A\:[\:(lq)^2(1+s_{+}s_{-})-(4m_t^2s-s^2)(1-s_{+}s_{-})
\non\\
&&\ \ \ \lspace -2(s-4m_t^2)(ls_{+}\;ls_{-}-Ps_{+}\;Ps_{-})
           -\:2\,lq(ls_{+}\;Ps_{-}-ls_{-}\;Ps_{+})\:]
\non\\
&&\ \ \ 
-4\:{\rm Re}(D_{V\!\!A})\:m_t\,[\:s(\psp-\psm)+lq(\lsp+\lsm)\:]\non\\
&&\ \ \ +2\:{\rm Im}(D_{V\!\!A})\:[\:lq\,\eps(\splus,\smin,q,l)
+\lsm\eps(\splus,P,q,l)+\lsp\eps(\smin,P,q,l)\:]
\non\\
&&\ \ \ +4\:E_V\:\m s(\lsp+\lsm)+4\:E_A\:\m\,lq(\psp-\psm)
\non\\
&&\ \ \ 
+4\:{\rm Re}(E_{V\!\!A})\:[\:2\mq(\lsp\;\psm-\lsm\;\psp)-lq\:s\:]
\non\\
&&\ \ \ +4\:{\rm Im}(E_{V\!\!A})\:\m[\:\eps(\splus,P,q,l)
+\eps(\smin,P,q,l)\:]
\non\\
&&\ \ \ -\:{\rm Re}(F_1)\:\mr[\:lq\;s(\lsp-\lsm)
-\{(lq)^2+4\mq s\}(\psp+\psm)\:]
\non\\
&&\ \ \ +2\:{\rm Im}(F_1)\:[\:s\,\eps(\splus,\smin,P,q)
+lq\,\eps(\splus,\smin,P,l)\:]
\non\\
&&\ \ \ +2\:{\rm Re}(F_2)\:s(\psp\;\lsm+\psm\;\lsp)
\non\\
&&\ \ \ -\:{\rm Im}(F_2)\:\frac{s}{m_t}
[\:\eps(\splus,P,q,l)-\eps(\smin,P,q,l)\:]
\non\\
&&\ \ \ -2\:{\rm Re}(F_3)\:lq(\psp\;\lsm+\psm\;\lsp)
\non\\
&&\ \ \ +\:{\rm Im}(F_3)\:\frac{lq}{m_t}
[\:\eps(\splus,P,q,l)-\eps(\smin,P,q,l)\:]
\non\\
&&\ \ \ -\:{\rm Re}(F_4)\:\frac{s}{m_t}
[\:lq\;(\psp+\psm)-(s-4\mq)(\lsp-\lsm)\:]
\non\\
&&\ \ \ -2\:{\rm Im}(F_4)\:
[\:\psp\eps(\smin,P,q,l)+\psm\eps(\splus,P,q,l)\:]
\non\\
&&\ \ \ +2\:{\rm Re}(G_1)\:
[\:\{ 4\mq s+(lq)^2-s^2 \}(1-\sss)-2s\,\psp\psm
\non\\
&&\ \ \ \lspace +lq(\lsp\;\psm-\lsm\;\psp)\:]
\non\\
&&\ \ \ -\:{\rm Im}(G_1)\:\frac{lq}{m_t}
[\:\eps(\splus,P,q,l)+\eps(\smin,P,q,l)\:]
\non\\
&&\ \ \ -\:{\rm Re}(G_2)\:\frac{s}{m_t}
[\:(s-4\mq)(\lsp+\lsm)-lq\;(\psp-\psm)\:]
\non\\
&&\ \ \ -2\:{\rm Im}(G_2)\:
[\:\psp\eps(\smin,P,q,l)-\psm\eps(\splus,P,q,l)\:]
\non\\
&&\ \ \ -\:{\rm Re}(G_3)\:\frac{lq}{m_t}
[\:lq\;(\psp-\psm)-(s-4\mq)(\lsp+\lsm)\:]
\non\\
&&\ \ \ -2\:{\rm Im}(G_3)\:lq\,\eps(\splus,\smin,q,l)
\non\\
&&\ \ \ +2\:{\rm Re}(G_4)\:
[\:(s-4\mq)(\psp\;\lsm-\psm\;\lsp)+2\,lq\;\psp\psm\:]
\non\\
&&\ \ \ +\:{\rm Im}(G_4)\:\mr (s-4\mq)[\:\eps(\splus,P,q,l)
+\eps(\smin,P,q,l)\:]\:\:\Bigr],
\label{distribution}
\end{eqnarray}
where
$$
P\equiv p_e+p_{\bar{e}}\:(=p_t+p_{\bar{t}})\,,\ \ l\equiv
p_e-p_{\bar{e}}\,,\ \ q\equiv p_t-p_{\bar{t}}\,,
$$
the symbol $\epsilon(a,b,c,d)$ means $\epsilon_{\mu\nu\rho\sigma}
a^\mu b^\nu c^\rho d^\sigma$ for $\epsilon_{0123}=+1$,
\begin{eqnarray}
&& D_V\:\equiv \:C\,[\,\avs-2\av\az\ci+\azs\cz+2(\av-\az\ci)
{\rm Re}(\delta\!A_\gamma)
\non\\
&& \lspace-2\{ \av\ci-\az\cz \}{\rm Re}(\delta\!A_Z)\,],
\non\\
&& D_A\:\equiv \:C\,[\,B_Z^2(1+v_e^2)\cci^2
-2B_Z v_e \cci{\rm Re}(\delta\!B_\gamma)
+2B_Z(1+v_e^2)\cci^2{\rm Re}(\delta\!B_Z)\,],
\non\\
&& D_{V\!\!A}\:\equiv \:C\,
[\,-\av\bz\ci+\az\bz\cz-\bz\ci(\delta\!A_\gamma)^*
\non\\
&& \lspace
+(A_\gamma -v_e \cci A_Z)\delta\!B_\gamma +\bz\cz(\delta\!A_Z)^* 
\non\\
&& \lspace-\{ \av\ci-\az\cz \}\delta\!B_Z \,],
\non\\
&& E_V\:\equiv \:2C\,
[\,\av\az\cci-\azs\ccz+\az\cci{\rm Re}(\delta\!A_\gamma)
+(\av\cci-2\az\ccz){\rm Re}(\delta\!A_Z) \,],
\non\\
&& E_A\:\equiv \:2C\,[\,-\bzs\ccz+B_Z \cci{\rm Re}(\delta\!B_\gamma)
-2\bz\ccz{\rm Re}(\delta\!B_Z) \,],
\non\\
&& E_{V\!\!A}\:\equiv \:C\,
[\,\av\bz\cci-2\az\bz\ccz+\bz\cci(\delta\!A_\gamma)^*
+A_Z \cci\delta\!B_\gamma
\non\\
&& \lspace-2\bz\ccz(\delta\!A_Z)^*
+(\av\cci-2\az\ccz)\delta\!B_Z \,],
\non\\
&& F_1\:\equiv \:C\,[\, -(\av-\az\ci)\delta\!D_\gamma
     +\{\av\ci-\az\cz\}\delta\!D_Z \,],
\non\\
&& F_2\:\equiv \:C\,[\, -\az\cci\delta\!D_\gamma
     -(\av\cci-2\az\ccz)\delta\!D_Z \,],
\non\\
&& F_3\:\equiv \:C\,[\, \bz\ci\delta\!D_\gamma-\bz\cz\delta\!D_Z \,],
\non\\
&& F_4\:\equiv \:C\,
[\, -\bz\cci\delta\!D_\gamma+2\bz\ccz\delta\!D_Z \,],
\non\\
&& G_1\:\equiv \:C\,[\, (\av-\az\ci)\delta C_\gamma
     -\{\av\ci-\az\cz\}\delta C_Z \,],
\non\\
&& G_2\:\equiv \:C\,[\, \az\cci\delta C_\gamma
     +(\av\cci-2\az\ccz)\delta C_Z \,],
\non\\
&& G_3\:\equiv \:C\,[\, -\bz\ci\delta C_\gamma +\bz\cz\delta C_Z \,],
\non\\
&& G_4\:\equiv \:C\,
[\, \bz\cci\delta C_\gamma-2\bz\ccz\delta C_Z \,],
\end{eqnarray}
and
$$
C \equiv 1/(4\sin^2\theta_W),\ \ \ \ \ \cci\equiv d \cdot 4\sw\cw.
$$
$D^{(0)}_{V,A,V\!\!A}$ used in \S$\:$2 (eq.{\ref{DVDA}) are the 
standard-model parts of the above $D_{V,A,V\!\!A}$. $D_{V,A,V\!\!A}$
and $E_{V,A,V\!\!A}$ are both defined the same way as in
ref.\cite{AS}, however $F_i$ ($i=1\sim 4$) differ by a factor
$-2im_t$ (i.e., $our\ F_i=-2im_t \times their\ F_i$). Finally it
should be emphasized that only linear terms in the non-standard
couplings have been kept.

\sec{Non-standard effects in $\mib{t}$ and $\bar{\mib{t}}$ decays}

We shall next consider non-standard effects in $t$ and $\bar{t}$
decays. We will adopt the following parameterization of the $Wtb$
vertex suitable for the $\twbdec$ and $\tbwbdec$ decays for the
on-shell $W$ :
\begin{eqnarray}
&&\!\!{\mit\Gamma}^{\mu}_{Wtb}=-{g\over\sqrt{2}}V_{tb}\:
\bar{u}(p_b)\biggl[\,\gamma^{\mu}(f_1^L P_L +f_1^R P_R)
-{{i\sigma^{\mu\nu}k_{\nu}}\over M_W}
(f_2^L P_L +f_2^R P_R)\,\biggr]u(p_t),\ \ \ \ \ \ \label{ffdef}\\
&&\!\!\bar{\mit\Gamma}^{\mu}_{Wtb}=-{g\over\sqrt{2}}V_{tb}^*\:
\bar{v}(p_{\bar{t}})
\biggl[\,\gamma^{\mu}(\bar{f}_1^L P_L +\bar{f}_1^R P_R)
-{{i\sigma^{\mu\nu}k_{\nu}}\over M_W}
(\bar{f}_2^L P_L +\bar{f}_2^R P_R)\,\biggr]v(p_{\bar{b}}),
\label{ffbdef}
\end{eqnarray}
where $P_{L/R}=(1\mp\gamma_5)/2$, $V_{tb}$ is the $(tb)$ element of
the Kobayashi-Maskawa matrix and $k$ is the momentum of $W$. Because
$W$ is on shell,\footnote{Note that we use the narrow-width
    approximation also for the $W$ propagator as mentioned in \S$
    \:$2.}\ 
the two additional form factors do not contribute. We show in the
appendix how dimension 6 operators affect these form factors. 

It is worth to know that the form factors for top and anti-top
satisfy the following relations \cite{cprelation}:
\begin{equation}
f_1^{L,R}=\pm\bar{f}_1^{L,R},\lspace f_2^{L,R}=\pm\bar{f}_2^{R,L},
\label{cprel}
\end{equation}
where upper (lower) signs are those for $C\!P$-conserving
(-violating) contributions.\footnote{Assuming $C\!P$-conserving
    Kobayashi-Maskawa matrix.}

$W\ell\nu$ couplings are treated within the SM as mentioned before:
\bea
&&{\mit\Gamma}^\mu_{W\ell\nu}=-\frac{g}{2\sqrt{2}}\,
  \bar{u}(p_\nu)\gamma^\mu(1-\gamma_5)v(p_{\ell^+}), \\
&&\bar{\mit\Gamma}^\mu_{W\ell\nu}=-\frac{g}{2\sqrt{2}}\,
  \bar{u}(p_{\ell^-})\gamma^\mu(1-\gamma_5)v(p_{\bar{\nu}}).
\eea

Assuming that
$\stackrel{\scriptscriptstyle(-)}{f_1^L}-1$,
$\stackrel{\scriptscriptstyle(-)}{f_1^R}$,
$\stackrel{\scriptscriptstyle(-)}{f_2^L}$ and
$\stackrel{\scriptscriptstyle(-)}{f_2^R}$ are small and keeping only
their linear terms, we obtain for the double differential spectrum in
$x$ and $\omega$ the following result:
\begin{equation}
\frac{1}{{\mit\Gamma}_t}
\frac{d^2{\mit\Gamma}_\ell}{dxd\omega}(t\to b\ell^+ \nu)
=\frac{1+\beta}{\beta}\;\frac{3 B_\ell}{W}
\omega\left[1+2{\rm Re}(f_2^R)\sqrt{r}\left(\frac{1}{1-\omega}-
\frac{3}{1+2r} \right)\right]. \label{t-decay}
\end{equation}
An analogous formula for $\bar{t}\to\bar{b}\ell^-\bar{\nu}$ holds
with $f_2^R$ replaced by $\bar{f}_2^L$ :
\begin{equation}
\frac{1}{{\mit\Gamma}_t}
\frac{d^2{\mit\Gamma}_\ell}{dxd\omega}(\bar{t}\to \bar{b}\ell^-
\bar{\nu})=\frac{1+\beta}{\beta}\;\frac{3 B_\ell}{W}\omega\left[
1+2{\rm Re}(\bar{f}_2^L)\sqrt{r}\left(\frac{1}{1-\omega}-
\frac{3}{1+2r} \right)\right]. \label{tbar-decay}
\end{equation}

Here let us give some comments on the $C\!P$-violating parameters.
Combining the above results with eq.(\ref{cprel}), we find that $C\!
P$-violating quantities in the decay processes are proportional to
${\rm Re}(f_2^R-\bar{f}_2^L)$ within our approximation. As will be
found in the appendix (eqs.(\ref{f2R}) and (\ref{fbar2L})), ${\rm Re}
(f_2^R-\bar{f}_2^L)$ becomes zero in the effective lagrangian
approach where only the SM particles are taken into account. Indeed
$C\!PT$ symmetry also requires this. That is, $C\!PT$ tells us that
order by order in perturbation expansion ${\mit\Gamma}(t \to all)=
{\mit\Gamma}(\bar{t}\to all)$. Then, if we assume only the standard
top-quark decay channel $t\to b W^+$ at the lowest order, one is led
to ${\rm Re}(f_2^R-\bar{f}_2^L)=0$ through the above (\ref{t-decay})
and (\ref{tbar-decay}). Non-zero ${\rm Re}(f_2^R-\bar{f}_2^L)$ may
emerge at the one-loop level, e.g., in SUSY with light neutralinos
and stops such that appropriate absorptive parts of vertex
corrections appear. This means observing non-zero ${\rm Re}(f_2^R-
\bar{f}_2^L)$ would be an evidence of not only some new interactions
but also some new relatively light particles.

\sec{The lepton-energy spectrum: non-standard results.}

Combining the results of the previous sections, we obtain the single
lepton-energy spectrum for $\epem \to \ell^\pm + \cdots$ :
\beq
\frac{1}{B_{\ell} \sigma_{e \bar{e} \to t \bar{t}} }
{\frac{d\sigma}{dx}}^{\!\pm}
=\sum_{i=1}^{3}c_i^\pm f_i(x),
\label{single}
\eeq
where $\pm$ corresponds to $\ell^{\pm}$,
$$
c_1^\pm =1,
$$
$$
c_2^\pm =
\lambda_1{\rm Re}(G_1)+\lambda_2\rdav
+\lambda_3{\rm Re}(\delta\!B_\gamma)
+\lambda_4\rdaz+ \lambda_5\rdbz \mp \xi,
$$
$$
c_3^+={\rm Re}(f_2^R),\ \ c_3^-={\rm Re}(\bar{f}_2^L),
$$
and
\beq
f_1(x)=f(x)+\eta\:g(x),\ \
f_2(x)=g(x),\ \
f_3(x)=\delta\!f(x)+\eta\:\delta g(x).
\label{fx}
\eeq
$\xi$ is the $C\!P$-violating parameter$\,$\footnote{Similarly to
    ${\rm Re}(f_2^R-\bar{f}_2^L)$, it will be seen in the appendix
    that within the effective lagrangian scenario $\xi=0$
    (eqs.(\ref{D-gamma}) and (\ref{D-Z})). Non-zero $\xi$ would also
    be generated through absorptive parts of loop diagrams involving
    relatively light non-standard particles.}\ 
in the production process used in refs. \cite{AS,GH_npb}
$$
\xi\equiv 2\,{\rm Re}(F_1)\,a_{V\!\!A},
$$
while $\lambda_i$ are defined as
\begin{eqnarray*}
&&\lambda_1 \equiv 2\,\beta^2 \eta \,a_{V\!\!A},
\\
&&\lambda_2 \equiv
-2\,a_{V\!\!A}\,[\,\eta (3-\beta^2)(\av-\az\ci)+2\bz\ci\,],
\\
&&\lambda_3 \equiv
4\,a_{V\!\!A}\,[\,A_\gamma -v_e \cci(A_Z -\eta\beta^2 B_Z)\,]
\\
&&\lambda_4 \equiv
2\,a_{V\!\!A}\,[\,\eta(3-\beta^2)\{\av\ci-\az\cz\}+2\bz\cz\,],
\\
&&\lambda_5 \equiv -4\,a_{V\!\!A}\,
[\,\av\ci -\cz(\az-\eta\beta^2\bz) \,].
\end{eqnarray*}
( $a_{V\!\!A}$ was defined just before eq.(\ref{S0}) in \S$\:$2.)
For the present input parameters, they are
\[
\lambda_1=\,0.1586,\ \ \lambda_2= -0.4303,\ \
\lambda_3=\,1.9578,\ \ \lambda_4=\,0.5635,\ \
\lambda_5=\,0.2684.
\]
$F_1$ and $G_1$ in $c_2^\pm$ are numerically related to the non-SM
parameters as
\[
\left(\begin{array}{c} F_1 \\ G_1 \end{array}\right)
=0.7035\left(\begin{array}{c} -\ddv \\ \dcv \end{array}\right)
+0.1205\left(\begin{array}{c} -\ddz \\ \dcz \end{array}\right).
\]
$\delta\!f(x)$ and $\delta g(x)$ are the functions derived in our
previous work, and their explicit form could be found in the appendix
of ref.\cite{GH_npb}. The functions $f_i(x)$ are shown in fig.
\ref{plot2d}.
\begin{figure}[t]
\postscript{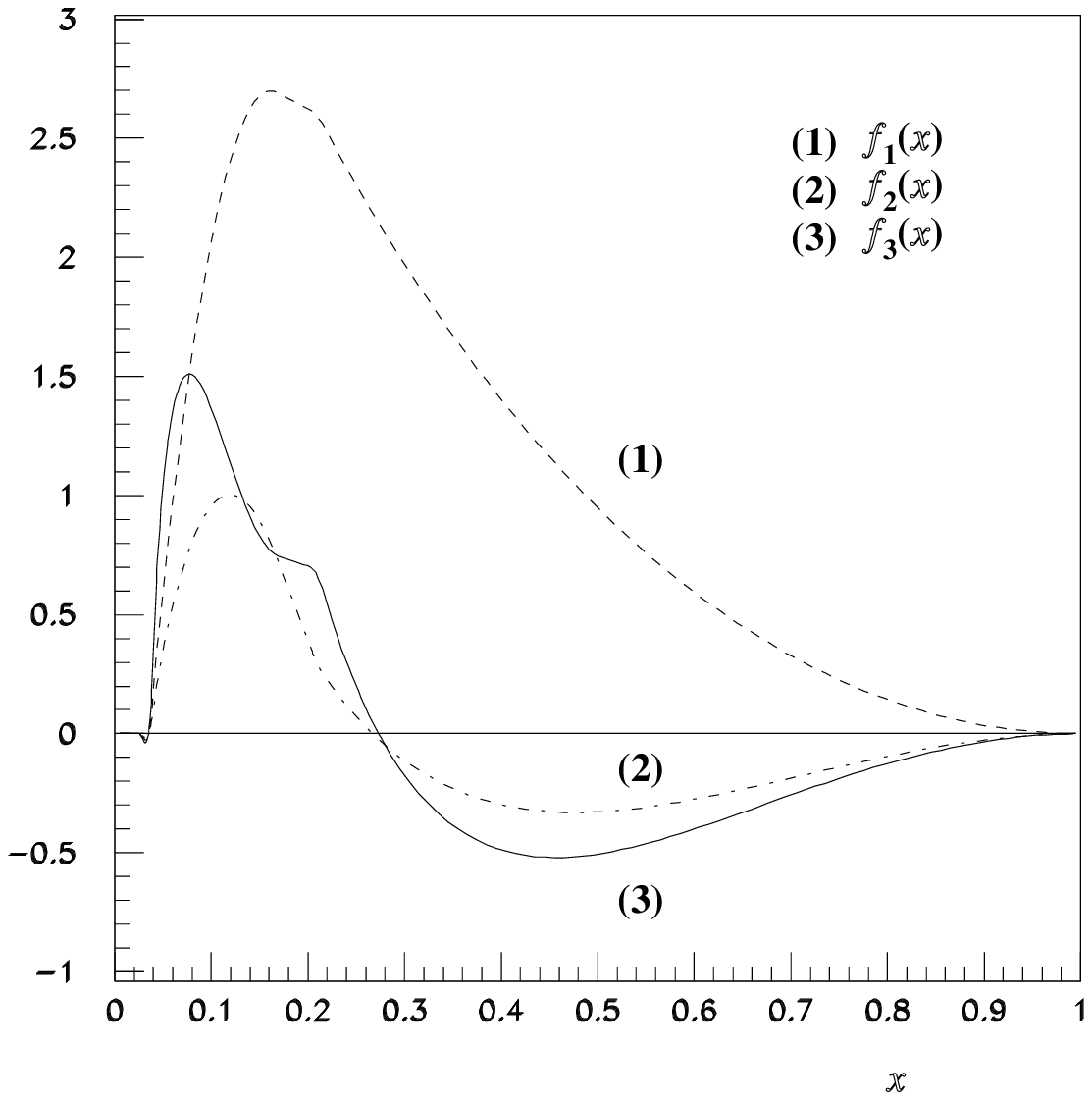}{0.80}
\vspace*{-0.5cm}
\caption{The functions $f_i(x)$ defined in eq.~(\protect\ref{fx}).}
\label{plot2d}
\end{figure}

The double lepton-energy spectrum is given by the following formula:
\beq
\frac{1}{B_{\ell}^{2} \sigma_{e \bar{e} \to t \bar{t}} } 
\frac{d^2 \sigma}{dx d\bar{x}}\;=\;\sum_{i=1}^{6}c_i f_i(x,\bar{x}),
\label{double}
\eeq
where
\[
c_1=1,\;\;c_2=\xi,\;\;c_3=\frac12{\rm Re}(f_2^R-\bar{f}_2^L),
\]
\[
c_4=\lambda_1'{\rm Re}(G_1)
+\lambda_2'\rdav+\lambda_3'{\rm Re}(\delta\!B_\gamma)
+\lambda_4'\rdaz+\lambda_5'\rdbz,
\]
\[
c_5=\lambda_1{\rm Re}(G_1)
+\lambda_2\rdav+\lambda_3{\rm Re}(\delta\!B_\gamma)
+\lambda_4\rdaz+\lambda_5\rdbz,
\]
\[
c_6=\frac12{\rm Re}(f_2^R+\bar{f}_2^L),
\]
and
\begin{eqnarray}
f_1(x,\bar{x})&\!\!=&\!\!f(x)f(\bar{x})+\eta\:[\:f(x)g(\bar{x})
   +g(x)f(\bar{x})\:]+\eta'g(x)g(\bar{x}), \non \\ 
f_2(x,\bar{x})&\!\!=&\!\!f(x)g(\bar{x})-g(x)f(\bar{x}), \non \\ 
f_3(x,\bar{x})&\!\!=&\!\!\delta\!f(x)f(\bar{x})
   -f(x)\delta\!f(\bar{x})\non \\ 
&\!\!+&\!\!\eta\:[\:\delta\!f(x)g(\bar{x})-f(x)\delta g(\bar{x})
   +\delta g(x)f(\bar{x})-g(x)\delta\!f(\bar{x})\:] \non \\
&\!\!+&\!\!\eta'[\:\delta g(x)g(\bar{x})-g(x)\delta g(\bar{x})\:],
   \non \\ 
f_4(x,\bar{x})&\!\!=&\!\!g(x)g(\bar{x}), \label{fxxbar} \non \\
f_5(x,\bar{x})&\!\!=&\!\!f(x)g(\bar{x})+g(x)f(\bar{x}), \non \\
f_6(x,\bar{x})&\!\!=&\!\!\delta\!f(x)f(\bar{x})
   +f(x)\delta\!f(\bar{x}) \non \\ 
&\!\!+&\!\!\eta\:[\:\delta\!f(x)g(\bar{x})+f(x)\delta g(\bar{x}) 
   +\delta g(x)f(\bar{x})+g(x)\delta\!f(\bar{x})\:] \non \\
&\!\!+&\!\!\eta'[\:\delta g(x)g(\bar{x})+g(x)\delta g(\bar{x})\:].
\end{eqnarray}
$\lambda_i'$ are defined similarly to $\lambda_i$ :
\begin{eqnarray*}
&&\lambda_1' \equiv 2(1+\beta^2\eta')\,a_{V\!\!A},
\\
&&\lambda_2' \equiv 2\beta^{-2}[\,1+\beta^2-\eta'\beta^2(3-\beta^2)
\,]\,a_{V\!\!A}\,(\,\av-\az\ci\,),
\\
&&\lambda_3' \equiv -4(1-\eta'\beta^2)\,a_{V\!\!A}\,B_Z v_e \cci,
\\
&&\lambda_4' \equiv 2\beta^{-2}[\,1+\beta^2-\eta'\beta^2(3-\beta^2)
\,]\,a_{V\!\!A}\,[\,-\av\ci+\az\cz \,],
\\
&&\lambda_5' \equiv 4(1-\eta'\beta^2)\,a_{V\!\!A}\,\bz\cz,
\end{eqnarray*}
and their values are
\[
\lambda_1'=\,2.4814,\ \ \lambda_2'= -0.1943,\ \
\lambda_3'=\,0.0278,\ \ \lambda_4'= -0.0333,\ \ \lambda_5'=\,0.2313.
\]
The functions $f_i(x,\bar{x})$ are plotted in fig. \ref{plots}.

\newpage
\begin{figure}[h]
\vspace*{-0.5cm}
\postscript{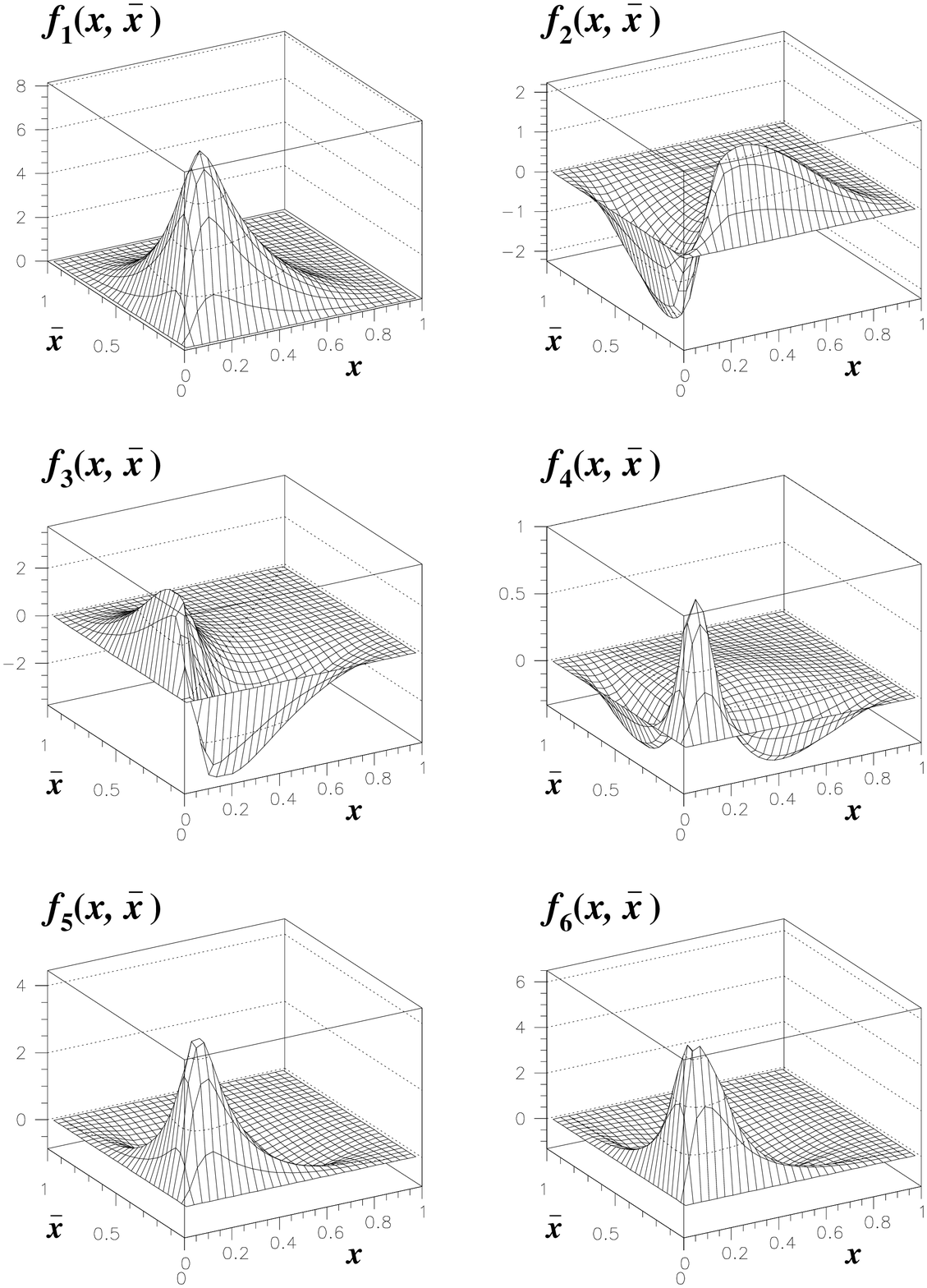}{0.77}
\caption{The functions $f_i(x,\bar{x})$ defined in
eq.~(\protect\ref{fxxbar}).}
\label{plots}
\end{figure}

\sec{The optimal observables}

Let us briefly summarize the main points of the optimal-observables 
technique~\cite{optimalization}. Suppose we have a cross section
$$
\frac{d\sigma}{d\phi}(\equiv{\mit\Sigma}(\phi))=\sum_i c_i f_i(\phi)
$$
where $f_i(\phi)$ are known functions of the location in final-state
phase space $\phi$ and $c_i$'s are model-dependent coefficients. The
goal would be to determine  $c_i$'s. It can be done by using
appropriate weighting functions $w_i(\phi)$ such that $\int w_i(\phi)
{\mit\Sigma}(\phi)d\phi=c_i$. Generally, different choices for
$w_i(\phi)$ are possible, but there is a unique choice such that the
resultant statistical error is minimized. Such functions are given by
\begin{equation}
w_i(\phi)=\sum_j X_{ij}f_j(\phi)/{\mit\Sigma}(\phi)\,, \label{X_def}
\end{equation}
where $X_{ij}$ is the inverse matrix of $M_{ij}$ which is defined as
\begin{equation}
M_{ij}\equiv \int {f_i(\phi)f_j(\phi)\over{\mit\Sigma}(\phi)}d\phi\,.
\label{M_def}
\end{equation}
When we take these weighting functions, the statistical uncertainty
of $c_i$ becomes
\begin{equation}
{\mit\Delta}c_i=\sqrt{X_{ii}\,\sigma_T/N}\,, \label{delc_i}
\end{equation}
where $\sigma_T\equiv\int (d\sigma/d\phi) d\phi$ and $N$ is the total
number of events.

\vskip 0.4cm \noindent
{\bf Single distribution}

Let us consider first the single-lepton inclusive process:
\beq
\epem \ra \ttbar \ra \ell^\pm \cdots,
\eeq
where $l=e$ or $\mu$ and dots stand either for jets or leptons. The
$M$ and $X$ matrices obtained from eq.(\ref{single}) are the same as
presented in ref.\cite{GH_plb} (the numerical values are a bit
different since the input data are not the same):
\[ (M_{ij})\;=\;\left[
\begin{array}{rrr}
1&    0   &   0     \\
0& 0.099  & 0.15 \\
0& 0.15   & 0.24  \\
\end{array}
\right],
\]
\[ (X_{ij})\;=\;\left[
\begin{array}{rrr}
1&    0    &   0      \\
0& 128.79  & -79.30   \\
0& -79.30  &  52.99   \\
\end{array}
\right].
\]
Since $M$ and $X$ are the same for $\ell^+$ and $\ell^-$ we obtain
the following statistical errors:
\beq
{\mit\Delta}c_2^\pm=11.35/\sqrt{N_{\ell}}\:,
\lspace {\mit\Delta}c_3^\pm =7.28/\sqrt{N_{\ell}}\:, 
\eeq
where $N_{\ell}$ is the expected number of detected single-lepton
events. This quantity is obtained from the integrated luminosity $L$
and lepton-tagging efficiency $\epsilon_{\ell}$ as $N_{\ell}=B_{\ell}
\sigma_{e \bar{e}\to t\bar{t}} L_{\rm eff}^{\ell}=127.9\,
L_{\rm eff}^{\ell}$, where $L_{\rm eff}^{\ell} \equiv \epsilon_{\ell}
L$ (in $\fbarn^{-1}$ units) and we estimated $\sigma_{e \bar{e} \to
t\bar{t}}$ to be $581.5\fbarn$ using $\alpha(s)=1/126$. Since
$\ell^+$ and $\ell^-$ events are statistically independent, we can
combine them when necessary. For example, we have ${\mit\Delta}\xi=
8.03/\sqrt{N_{\ell}}$ since $\xi=(c_2^- -c_2^+)/2$. Moreover, if we
were not interested in differences ($\cp$-violation) between the
$\ell^+$ and $\ell^-$ distributions but just concentrated on
determination of form factors  (the same for leptons and antileptons)
we could allow for both signs of leptons at the same time, which
would increase the branching-ratio-suppression factor from $18/81$ to
$36/81$ and increase expected statistics.\footnote{Here, the
    double-leptonic-decay modes were also included. In order to
    remove them one should replace the branching-ratio-suppression
    factors $18/81$ and $36/81$ by $12/81$ and $24/81$ respectively.} 

$3\sigma$ effects will be observable at $\sqrt{s}=500$ GeV if the
following relations are satisfied:
\beq
|c_2^\pm| \ge 3.01 \fbarn^{-1/2}/\sqrt{L_{\rm eff}^{\ell}}\:, \lspace
|c_3^\pm| \ge 1.93 \fbarn^{-1/2}/\sqrt{L_{\rm eff}^{\ell}}\:.
\eeq 
In table \ref{sing_lum} we show the square root of the effective
luminosity $\sqrt{L_{\rm eff}^{\ell}}$ obtained for some typical
representative values of $L$ and $\epsilon_{\ell}$ expected at
planned $\sqrt{s}=500\:\gev$ $\epem$ linear colliders
\cite{efficiency,JLC}.

\renewcommand{\arraystretch}{1.2}
\newcommand{\lw}[1]{\smash{\lower1.8ex\hbox{#1}}}
\begin{table}[h]
\bce 
\begin{tabular}{|c||c|c|c|c|}
\hline
\lw{$\epsilon_{\ell}$}&\multicolumn{4}{c|}{$L[\fbarn^{-1}]$}\\
\cline{2-5}
&20&40&100&200\\
\hline\hline
0.6&3.46&4.90&7.75&11.0\\
\hline
0.8&4.00&5.66&8.94&12.6\\
\hline
1.0&4.47&6.32&10.0&14.1\\
\hline
\end{tabular}\\
\vspace*{0.3cm}
\ece
\vspace*{-0.5cm}
\caption{Square root of the single-lepton effective luminosity: 
$\protect\sqrt{L_{\rm eff}^{\ell}}$.}
\label{sing_lum}
\end{table}

These are the most precise results which we can draw from the single
distribution (\ref{single}) alone. In order to get a higher
precision, therefore, it is important to combine it with other
independent information. For example, if we knew $c_3^\pm$ from some
other data and we only had to determine $c_2^\pm$ here, then the
corresponding matrices would become
\[ (M_{ij})\;=
\;\left[\begin{array}{rr} 1 & 0 \\ 0 & 0.099 \\ \end{array} \right],
\ \ \
   (X_{ij})\;=
\;\left[\begin{array}{rr} 1 & 0 \\ 0 & 10.10 \\ \end{array} \right],
\]
which leads to
\begin{equation}
{\mit\Delta}c_2^\pm=3.18/\sqrt{N_{\ell}}\:. \label{delc2}
\end{equation}
Similarly, if we already had $c_2^\pm$, then
\begin{equation}
{\mit\Delta}c_3^\pm=2.04/\sqrt{N_{\ell}}\:. \label{delc3}
\end{equation}
Discussion of the above results is left to the next section.
\vskip 0.4cm \noindent
{\bf Double distribution}

The double-lepton spectrum eq.(\ref{double}) leads to the following
$M$ and $X$ :
\[ (M_{ij})\;=\;\left[
\begin{array}{rrrrrr}
1 &0&0&0&0&0 \\
0 &  0.23 & -0.34 &  0      &  0      &  0       \\
0 & -0.34 &  0.54 &  0      &  0      &  0       \\
0 &  0    &  0    &  0.011  & -0.0038 & -0.0029  \\
0 &  0    &  0    & -0.0038 &  0.18   &  0.26    \\
0 &  0    &  0    & -0.0029 &  0.26   &  0.44
\end{array}
\right],
\]
\[ (X_{ij})\;=\;\left[
\begin{array}{cccccc}
1&0&0&0&0&0 \\
0 & 42.27 & 26.09 &    0    &   0    &    0       \\
0 & 26.09 & 17.94 &    0    &   0    &    0       \\
0 &  0    &  0    &  96.88  &   9.16 &  -4.88     \\
0 &  0    &  0    &   9.16  &  47.42 & -28.55     \\
0 &  0    &  0    &  -4.88  & -28.55 &  19.49
\end{array}
\right].
\]

\noindent
The statistical errors for the determination of the coefficients
$c_i$ calculated according to the formula (\ref{delc_i}) for the
double-lepton spectrum are shown in table~\ref{double_errors}.
Table \ref{doub_lum} provides the square root of the effective
luminosity $\sqrt{L_{\rm eff}^{\ell\ell}}$ similarly to table
\ref{sing_lum}, where $L_{\rm eff}^{\ell\ell}\equiv \epsilon_{\ell}^2
L$.
\newpage
\renewcommand{\arraystretch}{1.4}
\begin{table}[h]
\vspace*{-0.4cm}
\bce 
\begin{tabular}{|c||c|c|c|c|c|}
\hline
$i$&2&3&4&5&6\\
\hline
${\mit\Delta}c_i \sqrt{N_{\ell\ell}}$&6.50&4.24&9.84&6.89&4.41\\
\hline
$|c_i|\sqrt{L_{\rm eff}^{\ell\ell}}$
                         &3.68&2.40&5.57&3.89&2.50\\
\hline
\end{tabular}\\
\vspace*{0.3cm}
\ece
\vspace*{-0.5cm}
\caption{Standard deviations ${\mit\Delta}c_i$ expected for
measurements of $c_i$ defined for the double energy spectrum by
eq.(\protect\ref{double}) could be read from the second row, where
$N_{\ell\ell}=B_{\ell}^2 \sigma_{e \bar{e}\to t \bar{t}}
L_{\rm eff}^{\ell\ell}=28.14\, L_{\rm eff}^{\ell\ell}$ denotes the
expected number of double-lepton events. $L_{\rm eff}^{\ell\ell}
\equiv \epsilon_{\ell}^2 L$ (in $\fbarn^{-1}$ units) stands for the
effective integrated luminosity. The last row shows the minimal value
for $|c_i|\protect\sqrt{L_{\rm eff}^{\ell\ell}}$ (in $\fbarn^{-1/2}$
units) necessary for an observation of $3\sigma$ effects at $\protect
\sqrt{s}=500$ GeV.}
\label{double_errors}
\end{table}
\renewcommand{\arraystretch}{1.2}
\begin{table}[h]
\vspace*{-0.4cm}
\bce 
\begin{tabular}{|c||c|c|c|c|}
\hline
\lw{$\epsilon_{\ell}$}&\multicolumn{4}{c|}{$L[\fbarn^{-1}]$}\\
\cline{2-5}
&20&40&100&200\\
\hline\hline
0.6&2.68&3.79&6.00&8.49\\
\hline
0.8&3.58&5.06&8.00&11.3\\
\hline
1.0&4.47&6.32&10.0&14.1\\
\hline
\end{tabular}\\
\vspace*{0.3cm}
\ece
\vspace*{-0.5cm}
\caption{Square root of the double-lepton effective luminosity:
$\protect\sqrt{L_{\rm eff}^{\ell\ell}}$.} 
\label{doub_lum}
\end{table}

Again we can find from the diagonal elements of $M_{ij}$ what
precision we can get when other information is available and we have
here only one undetermined parameter left.

\sec{Summary, discussion and comments}

Next-generation linear colliders of $e^+ e^-$, NLC, will provide a
cleanest environment for studying top-quark interactions. There, we
shall be able to perform detailed tests of the top-quark couplings to
the vector bosons and either confirm the SM simple
generation-repetition pattern or discover some non-standard
interactions. In this paper, assuming the most general
($\cp$-violating {\it and} $\cp$-conserving) couplings for $\gamma
\ttbar$, $Z \ttbar$ and $W tb$, we have calculated in a
model-independent way the single- and the double-leptonic spectra.
Then, the recently proposed optimal-observables technique
\cite{optimalization} have been adopted to determine non-standard
couplings both through the single- and double-leptonic-spectra
measurements. The effective luminosity necessary for an observation
of $3\sigma$ effects at $\sqrt{s}=500$ GeV for given values of
non-standard couplings have been found. It would be very interesting
if we observed non-standard couplings to be non-zero. In particular,
finding non-zero $C\!P$-violating parameters must be exciting since
in that case not only new interactions but also new relatively-light
particles are required as discussed in \S$\:$4 and 5.

The results we have presented are the most precise ones which could
be obtained from the single or double distribution alone. As pointed
out in the previous section, combining various independent data is
important to achieve a higher precision, for which we gave a simple
example there. Indeed there are several articles in which such
comprehensive analysis has been performed. In ref.\cite{frey} full
reconstruction of events was assumed, and helicity amplitudes for the
production and decay processes were adopted to construct a
likelihood. Then the likelihood has been investigated varying a
single form factor at a time. Therefore one can compare expected
precision for a measurement of form factors with our example
eqs.(\ref{delc2}, \ref{delc3}) for ${\mit\Delta}c_{2/3}^\pm$ while
$c_{3/2}^\pm$ is known. In spite of the fact that the method adopted
here requires just lepton-energy measurement, it is seen that the
obtained precision has not been substantially reduced (notice that in
ref.\cite{frey} integrated luminosity of $10\fbarn^{-1}$ has been
applied).

Ladinsky and Yuan discussed prospects for measurements of top-quark
non-standard form factors at future linear colliders \cite{yuan}.
Their analysis based on the top-quark angular distribution assumes
however that its decays do not involve any non-standard couplings.
They provide algorithms to determine top-quark momentum through its
decay products, where full reconstruction of events is needed as in
ref.\cite{frey}. Taking into account that we measure here only
lepton energy, the results are consistent even though our precision
is slightly lower. Similar analysis including some non-standard
effects in top-quark decays has been performed in ref.\cite{schmidt}.
The precision for measurements of non-standard couplings estimated
there is again at the level of a few per cents. 

Angular-energy distribution of charged leptons originating from the
top-quark decay measured in the top-quark rest frame is known to be
very sensitive to the top-quark spin direction \cite{jezabek}.
Therefore one can adopt it as analyzer of the top-quark polarization
vector. Since the angular distribution of polarized quarks,
eq.(\ref{distribution}), is sensitive to non-standard interactions,
it is understandable that the energy spectrum of leptons does in fact
carry information on non-standard couplings in the production
process. The angular-energy distribution measured in the top-quark
rest frame is of course also sensitive to non-standard interactions
in the decay. Therefore it has been advocated by Je\.zabek and K\"uhn
in ref.\cite{jezabek_kuhn} as a possible test of $V\pm A$ charged
current structure. Numerical analysis performed by Schmitt
\cite{schmitt} shows that the expected precision for right-handed
charged current should be about $2\%$ of the left-handed coupling.

Finally, let us give a brief comment on the effects of radiative
corrections. All the non-standard couplings considered here may be
generated at the multi-loop level within the SM. As it has been
already mentioned, $\cp$-violating couplings $\ddv$, $\ddz$ and ${\rm
Re}(f_2^R-\bar{f}_2^L)$ requires at least two loops of the SM, so
they are negligible. On the other hand, $\cp$-conserving couplings
$\delta\!A_{\gamma,Z}$, $\delta\!B_{\gamma,Z}$, $\delta C_{\gamma,Z}$ 
and ${\rm Re}(f_2^R+\bar{f}_2^L)$ could be generated already at the
one-loop level approximation of QCD. Therefore, in order to
disentangle non-standard top-quark interactions and such QCD effects
it is important to calculate and subtract the QCD contributions from
the lepton-energy spectrum, this is however beyond the scope of this
paper. One should however remember that the QCD corrections include
also an emission of real gluons and therefore not only form factors
would be corrected but also the structure of matrix
elements\footnote{One could also allow for modification of the
    standard $g\ttbar$ coupling.}\
and the phase space would be different. Consequently, optimal
observables would need to be modified. Ref.\cite{jezabek,QCD}
provides literature on QCD corrections to $\ttbar$ production and/or
decay at linear colliders.

\vspace*{0.6cm}
\centerline{ACKNOWLEDGMENTS}

\vspace*{0.3cm}
We are grateful to M. Tanimoto and S. Matsumoto for sending us useful
piece of information reported at XVIII International Symposium on
Lepton-Photon Interactions. This work is supported in part by the
State Committee for Scientific Research (Poland) under grant 2 P03B
180 09 and by Maria Sk\l odowska-Curie Joint Fund II (Poland-USA)
under grant MEN/NSF-96-252.

\vspace*{0.6cm}
\noindent
{\bf Appendix}

The form factors discussed in the text could be derived within the
framework of the effective lagrangian parameterizing non-standard
corrections to the SM. 

The effective lagrangian approach requires a choice of the low-energy
particle content. In this paper we assume that the SM correctly
describes all such excitations (including the Higgs
particle).\footnote{Other approaches can be followed, assuming, for
    example, an extended scalar sector or the complete absence of
    light physical scalars.}\ 
Thus we imagine that there is a scale ${\mit\Lambda}$, independent of
the Fermi scale, at which the new physics becomes apparent. Since the
SM is renormalizable and ${\mit\Lambda}$ is assumed to be large, the
decoupling theorem~\cite{Appelquist} is applicable and requires that
all new-physics effects be suppressed by inverse powers of ${\mit
\Lambda}$. All such effects are expressed in terms of a series of
local {\it gauge invariant} ($SU(2)_L \times U(1)$) operators of
canonical dimension $ > 4 $; the catalogue of such operators up to
dimension 6 is given in ref.\cite{bw} (there are no dimension 5
operators respecting the global and local symmetries of the SM).

Adopting the notation of Buchm\"uller and Wyler we are listing all
the dimension 6 operators contributing to $\gamma t\bar{t}$ and $Z
t\bar{t}$ vertices: 
$$
\begin{array}{lcllcl}
{\cal O}_{qW}&\!\!=&\!\!
    i\bar{q}\tau^i\gamma_{\mu}D_{\nu}qW^{i\mu\nu},\;\;\;&
{\cal O}_{qB}&\!\!=&\!\!i\bar{q}\gamma_{\mu}D_{\nu}qB^{\mu\nu},\\
{\cal O}_{uB}&\!\!=&\!\!
    i\bar{u}\gamma_{\mu}D_{\nu}uB^{\mu\nu},\;\;\;&
{\cal O}_{\phi q}^{(1)}&\!\!=&\!\!
    i(\phi^{\dagger}D_{\mu}\phi)(\bar{q}\gamma^{\mu}q),\\
{\cal O}_{\phi q}^{(3)}&\!\!=&\!\!
    i(\phi^{\dagger}D_{\mu}\tau^i\phi)(\bar{q}\gamma^{\mu}\tau^i q),
    \;\;\;&
{\cal O}_{\phi u}&\!\!=&\!\!
    i(\phi^{\dagger}D_{\mu}\phi)(\bar{u}\gamma^{\mu}u),\\
{\cal O}_{Du}&\!\!=&\!\!(\bar{q}D_{\mu}u)D^{\mu}\tilde{\phi},\;\;\;&
{\cal O}_{\bar{D}u}&\!\!=&\!\!
    (D_{\mu}\bar{q}u)D^{\mu}\tilde{\phi},\\
{\cal O}_{uW\phi}&\!\!=&\!\!
    (\bar{q}\sigma_{\mu\nu}\tau^i u)\tilde{\phi}W^{i\mu\nu},\;\;\;&
{\cal O}_{uB\phi}&\!\!=&\!\!
    (\bar{q}\sigma_{\mu\nu}u)\tilde{\phi}B^{\mu\nu},
\end{array}
$$
where $\tau^i$ is the Pauli matrices. Some of them also contribute to
$Wtb$ vertex. In addition, $Wtb$ vertex receives corrections from the
following dimension 6 operators:
$$
\begin{array}{lcllcl}
{\cal O}_{\phi\phi}
&\!\!=&\!\!i(\phi^{\dagger}\epsilon
D_{\mu}\phi)(\bar{u}\gamma^{\mu}d),\;\;\;&
{\cal O}_{dW\phi}&\!\!=&\!\!
(\bar{q}\sigma_{\mu\nu}\tau^id)\phi W^{i\mu\nu}, \\
{\cal O}_{Dd}&\!\!=&\!\!(\bar{q}D_{\mu}d)D^{\mu}\phi,\;\;\;&
{\cal O}_{\bar{D}d}&\!\!=&\!\!(D_{\mu}\bar{q}d)D^{\mu}\phi.
\end{array}
$$
Given the above list, the whole lagrangian is written as 
\begin{equation}
{\cal L}={\cal L}^{SM}+\frac{1}{{\mit\Lambda}^2}
\sum_i\left( \alpha_i {\cal O}_i +  \hbox{ h.c.} \right). \non
\end{equation}

We have used the following parameterization of $\gamma$ and $Z$
vertices relevant for the production process:
\begin{eqnarray}
{\mit\Gamma}_{vt\bar{t}}^{\mu}&\!\!=&\!\!
\frac{g}{2}\,\bar{u}(p_t)\,\Bigl[\,\gamma^\mu\{\:A_v+\delta\!A_v
-(B_v+\delta\!B_v)\gamma_5\:\} \non\\
&\!\!+&\!\!
\frac{(p_t-p_{\bar{t}})^\mu}{2m_t}(\delta C_v-\delta\!D_v\gamma_5)
+\frac{(p_t+p_{\bar{t}})^\mu}{2m_t}(\delta\!E_v-\delta\!F_v\gamma_5)
\,\Bigr]\,v(p_{\bar{t}})
\end{eqnarray}
($v=\gamma/Z$). Here we kept the terms of $\delta\!E_v$ and $\delta
\!F_v$ for the later discussion though they do not contribute to our
processes as mentioned in the main text. Direct calculation leads to
the following results for the non-standard contributions to the form
factors:$\,$\footnote{Renormalization of the gauge-boson fields have
been omitted for simplicity. For details, see ref.\cite{bw}.}
\begin{eqnarray}
\dav&\!\!=&\!\!\frac1{{\mit\Lambda}^2}\Bigl[\:\frac{s}{g}\bigl\{\cw
{\rm Im}(\alpha_{uB}+\alpha_{qB})+\sw{\rm Im}(\alpha_{qW})\bigr\}
\non\\ 
&&\ +\frac{8m_t v}{g}\bigl\{\cw {\rm Re}(\alpha_{uB\phi})
+\sw {\rm Re}(\alpha_{uW\phi})\bigr\}\:\Bigr],
\\
\daz&\!\!=&\!\!\frac1{{\mit\Lambda}^2}\Bigl[-
\frac{v^2}{\cw}{\rm Re}(\alpha_{\phi u}+\alpha_{\phi q}^{(1)}
-\alpha_{\phi q}^{(3)}) \non\\
&&-\
\frac{s}{g}\bigl\{\sw{\rm Im}(\alpha_{uB}+\alpha_{qB})
-\cw{\rm Im}(\alpha_{qW}) \bigr\} \non\\
&&-\ \frac{8m_t v}{g}\bigl\{\sw {\rm Re}(\alpha_{uB\phi})
-\cw {\rm Re}(\alpha_{uW\phi})\bigr\}\:\Bigr],
\\ 
\dbv&\!\!=&\!\!\frac1{{\mit\Lambda}^2}
\Bigl[\:\frac{s}{g}\bigl\{\sw{\rm Im}(\alpha_{qW})
+\cw{\rm Im}(\alpha_{qB}-\alpha_{uB}) \bigr\}\:\Bigr], \label{del-bg}
\\
\dbz&\!\!=&\!\!\frac1{{\mit\Lambda}^2}\Bigl[\:
\frac{v^2}{\cw} {\rm Re}(\alpha_{\phi u}
-\alpha_{\phi q}^{(1)}+\alpha_{\phi q}^{(3)}) \non\\  
&&+\ \frac{s}{g}\bigl\{\sw{\rm Im}(\alpha_{uB}-\alpha_{qB})
+\cw{\rm Im}(\alpha_{qW})\bigr\}\: \Bigr],
\\
\dcv&\!\!=&\!\!\frac1{{\mit\Lambda}^2}\Bigl[
-\frac{8m_t v}{g}\bigl\{\cw {\rm Re}(\alpha_{uB\phi})
+\sw {\rm Re}(\alpha_{uW\phi})\bigr\}\: \Bigr],
\\
\dcz&\!\!=&\!\!\frac1{{\mit\Lambda}^2}\Bigl[\:
\frac{8m_t v}{g}\bigl\{\sw {\rm Re}(\alpha_{uB\phi})
-\cw {\rm Re}(\alpha_{uW\phi})\bigr\} \non\\
&&-\
\frac{m_t v}{\cw}{\rm Re}(\alpha_{Du}-\alpha_{\bar{D}u})\:\Bigr],
\\
\ddv&\!\!=&\!\!\frac1{{\mit\Lambda}^2}\Bigl[\:
i\frac{4m_t^2}{g}\bigl\{\cw{\rm Re}(\alpha_{uB}-\alpha_{qB})
-\sw {\rm Re}(\alpha_{qW}) \bigr\}\non\\
&&+\ i\frac{8m_t v}{g}\bigl\{\cw {\rm Im}(\alpha_{uB\phi})
+\sw {\rm Im}(\alpha_{uW\phi})\bigr\}\: \Bigr], \label{D-gamma}
\\
\ddz&\!\!=&\!\!\frac1{{\mit\Lambda}^2}\Bigl[
-i\frac{4m_t^2}{g}\bigl\{\sw{\rm Re}(\alpha_{uB}-\alpha_{qB})
+\cw{\rm Re}(\alpha_{qW}) \bigr\} \non\\ 
&&-\ i\frac{8m_t v}{g}\bigl\{\sw {\rm Im}(\alpha_{uB\phi})
-\cw {\rm Im}(\alpha_{uW\phi})\bigr\} \non\\
&&+\
i\frac{m_t v}{\cw}{\rm Im}(\alpha_{Du}-\alpha_{\bar{D}u})\:\Bigr],
\label{D-Z}
\\
\dev&\!\!=&\!\!0, \label{del-eg}
\\
\dez&\!\!=&\!\!\frac1{{\mit\Lambda}^2}\Bigl[\:
i\frac{m_tv}{\cw}{\rm Im}(\alpha_{Du}+\alpha_{\bar{D}u})\:\Bigr],
\\
\dfv&\!\!=&\!\!\frac1{{\mit\Lambda}^2}\Bigl[
-\frac{4m_t^2}{g}\bigl\{\sw {\rm Im}(\alpha_{qW})+\cw{\rm
Im}(\alpha_{qB}-\alpha_{uB})\bigr\}\: \Bigr], \label{del-fg}
\\
\dfz&\!\!=&\!\!\frac1{{\mit\Lambda}^2}\Bigl[
-\frac{4m_t^2}{g}\bigl\{\cw {\rm Im}(\alpha_{qW})-\sw{\rm
Im}(\alpha_{qB}-\alpha_{uB})\bigr\} \non\\
&&-\
\frac{m_tv}{\cw}{\rm Re}(\alpha_{Du}+\alpha_{\bar{D}u}) \:\Bigr],
\end{eqnarray}
where the Higgs vacuum expectation value is $v$ (not $v/\sqrt{2}$) as
in ref.\cite{bw}.

Let us briefly discuss constraints on the general form of
${\mit\Gamma}^\mu_{\gamma t\bar{t}}$ from $U(1)_{\rm EM}$ symmetry.
This coupling, obtained as the matrix element of the EM current
$j^\mu(x)$ at $x=0$, is given as
\begin{eqnarray*}
\!\!\!\vev{p_t,p_{\bar{t}}|j^\mu(0)|0} 
&\!\!=&\!\!\bar{u}(p_t)\,
\Bigl[\,\gamma^\mu(A_\gamma+\dav-\dbv\gamma_5) 
+\frac{(p_t -p_{\bar{t}})^\mu}{2 m_t}(\dcv-\ddv\gamma_5)\\
&\!\!+&\!\!\frac{(p_t +p_{\bar{t}})^\mu}{2m_t}
(\dev-\dfv\gamma_5)\,\Bigr]\,v(p_{\bar{t}}).
\end{eqnarray*}
From the current conservation ($U(1)_{\rm EM}$ symmetry),
$\partial_\mu j^\mu(x)=0$,
\begin{eqnarray*}
&&0=\vev{p_t,p_{\bar{t}}|\partial_\mu j^\mu(0)|0}
=i(p_t +p_{\bar{t}})_\mu \vev{p_t,p_{\bar{t}}|j^\mu(0)|0} \\
&&\phantom{0}
=i\bar{u}(p_t)\,\Bigl[\,-2m_t \dbv\gamma_5
+\frac{s}{2 m_t}(\dev-\dfv\gamma_5)\,\Bigr]\,v(p_{\bar{t}}).
\end{eqnarray*}
As it is seen from above equation $U(1)_{\rm EM}$ gauge symmetry
requires the following relations to hold:
\begin{equation}
\dev=0, \;\;\;\;\;\; \dbv =-\frac{s}{4 m_t^2}\,\dfv.
\end{equation}
That is, {\it $U(1)_{EM}$ could be maintained even with non-zero
axial coupling $B_{\gamma}$}. Indeed, it could be directly checked
that the above relations are satisfied among eqs.(\ref{del-bg}),
(\ref{del-eg}) and (\ref{del-fg}). 

Using the notation defined by eqs.(\ref{ffdef}) and (\ref{ffbdef}),
the non-standard parts of the form factors contributing to the
top-quark decays read: 
\begin{eqnarray}
&&f^L_1=\frac1{V_{tb}{\mit\Lambda}^2}
  \Bigl[\:-\frac{2M_W^2}{g}\im(\alpha_{qW})
  +\frac{m_tv}{2}(\alpha_{Du}-\alpha_{\bar{D}u})
  -2v^2\alpha_{\phi q}^{(3)}\:\Bigr], \\
&&f^R_1=\frac1{V_{tb}{\mit\Lambda}^2}
  \Bigl[\:v^2\alpha_{\phi\phi}^*
  +\frac{m_tv}{2}(\alpha_{Dd}^*-\alpha_{\bar{D}d}^*)\:\Bigr], \\
&&f^L_2=\frac1{V_{tb}{\mit\Lambda}^2}
  \Bigl[\:-\frac{4M_Wv}{g}\alpha_{dW\phi}^*
  -\frac{M_Wv}{2}(\alpha_{Dd}^*-\alpha_{\bar{D}d}^*)\:\Bigr], \\
&&f^R_2=\frac1{V_{tb}{\mit\Lambda}^2}
  \Bigl[\:-\frac{4M_Wv}{g}\alpha_{uW\phi}
  -\frac{M_Wv}{2}(\alpha_{Du}-\alpha_{\bar{D}u})
  +i\frac{2m_tM_W}{g}\re(\alpha_{qW})\:\Bigr],~~~\label{f2R} \\
&&\bar{f}^L_1=\frac1{V_{tb}^*{\mit\Lambda}^2}
  \Bigl[\:-\frac{2M_W^2}{g}\im(\alpha_{qW})
  +\frac{m_tv}{2}(\alpha_{Du}^*-\alpha_{\bar{D}u}^*)
  -2v^2\alpha_{\phi q}^{(3)\,*}\:\Bigr], \\
&&\bar{f}^R_1=\frac1{V_{tb}^*{\mit\Lambda}^2}
  \Bigl[\:v^2\alpha_{\phi\phi}
  +\frac{m_tv}{2}(\alpha_{Dd}-\alpha_{\bar{D}d})\:\Bigr], \\
&&\bar{f}^L_2=\frac1{V_{tb}^*{\mit\Lambda}^2}
  \Bigl[\:-\frac{4M_Wv}{g}\alpha_{uW\phi}^*
  -\frac{M_Wv}{2}(\alpha_{Du}^*-\alpha_{\bar{D}u}^*)
  -i\frac{2m_tM_W}{g}\re(\alpha_{qW})\:\Bigr],~~~\label{fbar2L} \\
&&\bar{f}^R_2=\frac1{V_{tb}^*{\mit\Lambda}^2}
  \Bigl[\:-\frac{4M_Wv}{g}\alpha_{dW\phi}
  -\frac{M_Wv}{2}(\alpha_{Dd}-\alpha_{\bar{D}d})\:\Bigr].
\end{eqnarray}
Similar studies for the form factors have been recently performed in
ref.\cite{whisnant}. 

\vspace*{0.6cm}

\end{document}